\documentclass[nocopyrightspace,preprint,nonatbib,10pt]{sigplanconf}
\pdfoutput=1

\let\bibfont\relax

\expandafter\let\csname ver@natbib.sty\endcsname\relax

\usepackage[backend=bibtex,style=alphabetic]{biblatex}
\renewcommand*{\bibfont}{\small}

\title{ManyDSL: A Host for Many Languages}

\newcommand{\inst}[1]{\textsuperscript{#1}}
\authorinfo{Piotr Danilewski\inst{1,2,3} \and Philipp Slusallek\inst{1,2,4}}
{\inst{1}Saarland University, Germany
\and
\inst{2}Intel Visual Computing Institute, Germany
\and\\
\inst{3}Theoretical Computer Science, Jagiellonian University, Poland
\and\\
\inst{4}Deutsches Forschungszentrum f\"{u}r K\"{u}nstliche Intelligenz, Germany}
{piotr.danilewski@tcs.uj.edu.pl\hspace{5mm}slusallek@cg.uni-saarland.de\vspace{-.5cm}}

\usepackage{etex}
\usepackage[utf8]{inputenc}
\usepackage[T1]{fontenc}

\usepackage{amsfonts}

\usepackage{microtype}

\usepackage{acronym}
\usepackage{amsthm} 
\usepackage{amssymb}
\usepackage{amsmath}
\usepackage{mathtools}
\usepackage{multirow}
\usepackage{listings}
\usepackage{newfloat}
\usepackage{relsize}
\usepackage{verbatim}
\usepackage[labelfont=bf]{caption}
\usepackage[labelfont=bf,subrefformat=parens]{subcaption}
\usepackage{tikz}
\usepackage{ctable}
\usetikzlibrary{arrows,decorations,backgrounds,positioning,fit,petri}
\usetikzlibrary{shapes.multipart}
\usetikzlibrary{fadings}
\usetikzlibrary{patterns}

\ifx\pdfoutput\undefined
    \usepackage[final]{graphicx}
    \usepackage{hyperref}
\else
    \usepackage[pdftex]{hyperref}
    \hypersetup{pdftex,linkcolor=blue,citecolor=magenta,urlcolor=red,colorlinks=true}
\fi

\input{template/tikzshapes.tex}
\definecolor{darkgreen}{rgb}{0.2,0.7,0.1}
\definecolor{darkbronze}{rgb}{0.4,0.4,0.1}
\definecolor{darkgrey}{rgb}{0.3,0.3,0.3}
\lstdefinelanguage{CUDA}
[ISO]{C++}
{
morekeywords={
__global__,__host__,__device__,__constant__,__shared__,
gridDim,blockIdx,blockDim,threadIdx,
char1,char2,char3,char4,
uchar1,uchar2,uchar3,uchar4,
short1,short2,short3,short4,
ushort1,ushort2,ushort3,ushort4,
int1,int2,int3,int4,
uint1,uint2,uint3,uint4,
long1,long2,long3,long4,
ulong1,ulong2,ulong3,ulong4,
longlong1,longlong2,
float1,float2,float3,float4,
double1,double2,
dim1,dim2,dim3,dim4,
tex1Dfetch,tex1D,tex2D,tex3D,
__float_as_int,__int_as_float,
__float2int_rn,__float2int_rz,__float2int_ru,__float2int_rd,
__float2uint_rn,__float2uint_rz,__float2uint_ru,__float2uint_rd,
__int2float_rn,__int2float_rz,__int2float_ru,__int2float_rd,
__uint2float_rn,__uint2float_rz,__uint2float_ru,__uint2float_rd,
__fadd_rz,__fmul_rz,__fdividef,
__mul24,__umul24,__mulhi,__umulhi,__mul64hi,__umul64hi,
min,umin,fminf,fmin,max,umax,fmaxf,fmax,
abs,fabsf,fabs,rsqrtf,rsqrt,sqrtf,sqrt,__sinf,sinf,sin,__cosf,cosf,cos,
__sincosf,sincosf,sincos,__expf,expf,exp,__logf,logf,log,
__syncthreads,
__threadfence,
__threadfence_block
}
}

\lstdefinelanguage{pseudoCUDA}
[ISO]{C++}
{
morekeywords={
global,host,device,
function,
threadIdx,
threadCount,
blockIdx,
blockCount,
synchronize,
atomic,
in,
do
}
}

\lstdefinelanguage{Impala}
[ISO]{C++}
{
morekeywords={
def, kind, errorType, typeof
}
}

\definecolor{deepcpsstage}{rgb}{0.85,0.45,0.2}
\definecolor{deepcpspseudo}{rgb}{0,0.2,0.85}
\definecolor{langdslarg}{rgb}{0.2,0.55,0.75}
\definecolor{langdslargdefault}{rgb}{0.75,0.10,0.55}
\definecolor{langdsldeepcps}{rgb}{0,0.3,0.1}

\lstdefinelanguage{deepcps}
{
morekeywords={
fix, in, let, stage, always, never,
bool, int, float, fn, meta, function, grammar, epsilon
},
moredelim=[is][\color{deepcpsstage}\ttfamily]{'}{'},
moredelim=[is][\color{deepcpspseudo}\ttfamily\slshape]{"}{"},
moredelim=[is][\ttfamily\slshape]{↔}{↔},
moredelim=[is][\color{langdslarg}\ttfamily]{|}{|},
moredelim=[is][\color{langdsldeepcps}\ttfamily]{▼}{▼},
moredelim=[is][\color{langdslargdefault}\ttfamily]{§}{§},
literate={▲}{{\$}}1 {""}{{"}}1 {''}{{'}}1 {||}{{|}}1
}

\definecolor{langdslcode}{rgb}{0.10,0.25,0.65}

\lstdefinelanguage{langdsl}
{
frame=single,
morekeywords={
language, fragment, keyword, symbol, token, ignore arguments, rule, action, epsilon
},
moredelim=[is][\color{langdslcode}\ttfamily\slshape]{<}{>},
moredelim=[is][\color{deepcpspseudo}\ttfamily\slshape]{""}{""},
}

\definecolor{deepcpsstage}{rgb}{0.85,0.45,0.2}

\lstdefinelanguage{cstage}[ISO]{C++}
{
morekeywords={fn,stage,def,var},
moredelim=[is][\color{deepcpsstage}\ttfamily]{'}{'}
}

\lstdefinelanguage{MetaML}[]{ML}
{
moredelim=[is][\color{deepcpsstage}]{|}{|},
moredelim=[is][\bfseries\color{deepcpsstage}]{!}{!},
morekeywords={lift,run}
}

\lstdefinelanguage{P}
[]{Impala}
{
morekeywords={
thread, task, shared, solid
},
deletekeywords={global}
}

\lstdefinelanguage{PRT}
[]{P}
{
morekeywords={
persistent, closure, functor
}
}

\newcommand{\inline}[1]{\lstinline[basicstyle=\ttfamily]~#1~}

\newcommand{\ignore}[1]{}

\newcommand{\sfun}{\lambda\Stage{y}\overline{x} . b}
\newcommand{\AND}{\ \texttt{and}\ }
\newcommand{\OR}{\ \texttt{or}\ }
\newcommand{\NOT}{\texttt{not}\ }

\newcommand{\Stage}[1]{\texttt{[}#1\texttt{]}}

\newenvironment{itemize*}
{ \begin{list}{\labelitemii}{
    \setlength{\topsep}{0pt}
    \setlength{\parskip}{0pt}
   \setlength{\itemsep}{-3pt}
    \setlength{\leftmargin}{1em}
}}
{\end{list} }

\newenvironment{enumerate*}
{ \begin{list}{\arabic{enumi}.}{\usecounter{enumi}
    \setlength{\topsep}{0pt}
    \setlength{\parskip}{0pt}
   \setlength{\itemsep}{-1pt}
    \setlength{\leftmargin}{1.4em}
}}
{\end{list} }

\DeclareFloatingEnvironment[
    fileext=lop,
    name=Listing]{code}
\DeclareCaptionSubType{code}

\newcommand{\highlightcode}[6]{
\shade [left color=#1!30, middle color=white, shading=axis,shading angle=45] (#2,#3) rectangle (#4,#5);
\draw[color=#1](#2,#3) -- (#4,#3);
\draw[color=#1](#2,#3) -- (#2,#5);
\node[anchor=north east,inner sep=0.3] at (#4,#3) {\color{#1} \emph{#6}};
}
\newcommand{\highlightsolid}[5]{
\fill [color=#1] (#2,#3) rectangle (#4,#5);
\draw[color=#1](#2,#3) -- (#4,#3);
\draw[color=#1](#2,#3) -- (#2,#5);
}

\newacro{CPS}{Continuation Passing Style}
\newacro{LMS}{Lightweight Modular Staging}
\newacro{DSL}{Domain-Specific Language}

\newcommand{\includeComments}{%
\newcommand{\createComment}[3]{%
\expandafter \newcommand \csname
##1\endcsname[1]{{{\color{##3}\textsf{\textbf{[##2####1]}}}}}%
}}
\newcommand{\includeRemarks}{%
\newcommand{\createRemark}[3]{%
\expandafter \newcommand \csname
##1\endcsname{{{\color{##3}\textsf{\textbf{[##2]}}}}\xspace}%
}}
\newcommand{\noComments}{%
\newcommand{\createComment}[3]{%
\expandafter \newcommand \csname ##1\endcsname[1]{}%
}}

\newcommand{\noRemarks}{%
\newcommand{\createRemark}[3]{%

\expandafter \newcommand \csname ##1\endcsname[1]{}%
}}

\includeComments
\includeRemarks
\createComment{todo}{}{red}
\createRemark{tocite}{CITE}{red}

\hypersetup{linkcolor=black,citecolor=black,urlcolor=black,colorlinks=true}

\renewcommand{\P}[1]{\texttt{\scriptsize P\ref{lst:p#1}}}

\renewcommand{\topfraction}{.9}

\renewcommand{\floatsep}{8pt}
\renewcommand{\textfloatsep}{10pt}
\tikzfading[name=fade right,
  left color=transparent!0, right color=transparent!100]

\bibliography{literature}

\begin{document}

\maketitle
\begin{abstract}

Domain-specific languages are becoming increasingly important.
Almost every application touches multiple domains.
But how to define, use, and combine multiple DSLs within the same application?

The most common approach is to split the project along the domain boundaries into multiple pieces and files.
Each file is then compiled separately.
Alternatively, multiple languages can be embedded in a flexible host language:
within the same syntax a new domain semantic is provided.

In this paper we follow a less explored route of metamorphic languages.
These languages are able to modify their own syntax and semantics on the fly,
thus becoming a more flexible host for DSLs.

Our language allows for dynamic creation of grammars and switching languages where needed.
We achieve this through a novel concept of Syntax-Directed Execution.
A language grammar includes semantic actions that are pieces of functional code executed immediately during parsing.
By avoiding additional intermediate representation, connecting actions from different languages and domains is greatly simplified.
Still, actions can generate highly specialized code though lambda encapsulation and Dynamic Staging.

\end{abstract}


%


\section{Introduction}
\label{sec:introduction}

``Languages shape thought''. This Sapir-Whorf Hypothesis refers to human-speaking languages and the perception of the real world~\cite{Whorf, Sapir}.
This is no less true for programming languages~\cite{BlubParadox}.
A language style and the paradigms it supports can inadvertently put us in one mindset, preventing us from seeing alternative solutions~\cite{Dijkstra-Truths}.

\subsection{Domain Specific Languages}

For that reason, in an ideal situation each domain should have its own Domain Specific Language (DSL)
that best represents the concepts of the domain and guides the way of thinking of the programmer in the right direction.
However, creating such a language from scratch is not an easy task.
Available tools, such as YACC~\cite{yacc, LexYacc} or ANTLR~\cite{ANTLR} help define custom grammars
and perform basic operations on the created Abstract Syntax Tree (AST). 
However, more advanced aspects of language creation, such as variable lookup rules, type checking, or translation to other representation still needs to be performed by the language designer.

Languages created in such a way are independent one from another.
Yet, it is rare for a computer project to touch only a single domain.
Most applications deal with multiple domains, such as UI, database, communication, work scheduling, etc...

In order for a DSL to be used in practice, it needs to exchange data with other DSLs.
Such communication is often limited, inefficient, and unsafe:
external protocols, functions, or raw strings and files are used.




\begin{table*}
\begin{tabular}{ccl}
 Embedding & Host & Example \\
\hline
  & SQL &
\begin{lstlisting}	
SELECT Name, Surname FROM Members WHERE Age = 18 
\end{lstlisting} \\
\hline
Haskell/DB~\cite{EDSL-HaskellDB} & Haskell &
\begin{lstlisting}
do r <- table Members;
	 restrict ▲ r!Age .==. constant 18
	 project ▲ Name << r!Name # Surname << r!Surname
\end{lstlisting}
\\
\hline
LINQ~\cite{LINQ} & C\# &
\begin{lstlisting}
Members
    .Where(row => row.Age == 18).Select(row => new {row.Name, row.Surname});
\end{lstlisting}
\\
\hline
jOOQ~\cite{jOOQ} & Java &
\begin{lstlisting}
create.select(MEMBERS.NAME,MEMBERS.SURNAME)
    .from(MEMBERS).where(MEMBERS.AGE.eq(18)).fetch();
\end{lstlisting}
\\
\hline
Slick~\cite{Slick} & Scala &
\begin{lstlisting}
for {m <- Members if m.Age === 18} yield (m.Name, m.Surname)
\end{lstlisting}
\\
\end{tabular}
\caption{\label{table:sql_embedding}
Comparison of different SQL embeddings into general-purpose languages.
The same semantic meaning given in a syntax not specific to the domain makes the message harder to comprehend by an inexperienced programmer.
Note that the grammar of C\# features a construct specific to SQL which we do not show here -- we use only the general-purpose subset of C\#.}
\end{table*}

Inter-DSL communication is simpler when the DSL is \emph{embedded} within a more generic \emph{host language}~\cite{EmbedDSL}
through a set of overloaded functions, operators or macros.
Such languages are executed as part of the host code, and can exchange data between different languages through the host.
However, such DSLs are constrained by the syntax of the host, and cannot use their most natural and expected syntactic form.
In \autoref{table:sql_embedding} for example we show how the same SQL syntax is realized in various embedded database languages
 -- in all these cases, the obtained syntax is more complex and cluttered, hiding the original meaning of the code.

A \emph{metamorphic language} combines the strength of both stand-alone and embedded languages,
by allowing itself to be modified on the fly.
As a host it offers a common base connecting all DSLs while at the same time permits each language to have its own syntax and semantic.

%
%
%

\subsection{Contributions}

In this paper we present a metamorphic language designed with multi-DSL support in mind.
We combine the syntactic flexibility of stand-alone languages with the composability and inter-DSL communication of an embedded approach:
\begin{itemize}
\item We facilitate the dynamic creation of grammars. New grammars can be defined, combined, and used as a library.
\item We define a Language Programming Interface (LPI), defining how one DSL can be used by another.
The internal structure of a language does not impact the LPI.
\item Our implementation permits language switching and communication between different DSLs through the LPI.
\end{itemize}


\section{Related Work}
\label{sec:related}

\subsection{Parser Generators}
\label{ch:related:pargen}
A common tool for generating parsers, that now has many derivatives, is YACC~\cite{yacc}.
It is able to produce an LALR(1) parser~\cite{LRParsing} --- a practical approach to parsing of a subset of LR(1) grammars~\cite{LRKnuth}.
Further tools support more powerful classes of grammars: 
Generalized LR (GLR)~\cite{GLR}, packrat parsing~\cite{Packrat} or LL(*) parsing~\cite{ANTLR}.

While the classes of grammars change, the basic principle of specifying the language remains.
The description is given in a format similar to a Backus-Naur form~\cite{BNF},
augmented with \emph{attributes} and \emph{actions}. Each production is a concrete top-level entry and cannot be reused in another grammar.
The generated code is represented within the actions as text or as an AST that is being generated, or both.
These tools are typically used to produce stand-alone languages.
They provide no facilities for communication between DSLs or language switching.

\subsection{Parser Combinators}
\label{ch:parcomb}

Instead of using a dedicated tool, a parser can be defined in a functional language as a \emph{parser combinator}~\cite{Wadler,FuncParser,MonadParser}.
Each parser is a first-class entity that can be created dynamically and combined together to form bigger, more complex parsers.

Classical parser combinators create a parser that is highly redundant and may backtrack multiple times.
Although extra computation can be avoided through lazy evaluation and a careful design of the combinators,
in practice inefficient or even ambiguous parsers can be easily created by accident.
Many attempts have been made to address these issues~\cite{EffCombinator,Parsec}.
Furthermore, combinators can use staging to optimize themselves early, producing a simpler representation of the grammar that performs faster~\cite{ParserCombStaged}.

\subsection{Language Embedding}
Languages can be created by embedding them into a general-purpose host language, such as Haskell, ML, or Scala.
These DSLs use the same syntax as the host, but the names are overloaded to serve the new semantics.
Having the common base, DSLs can use it to relay information between the domains.
For this reason, this approach is particularly common when defining small, embedded DSLs (EDSLs).
For example, Haskell has been used extensively to define geometric operations~\cite{Experiment-Haskell},
COM component scripting~\cite{EDSL-COM-in-Haskell},
hardware design~\cite{EDSL-Lava}, or
server-side web scripting~\cite{EDSL-web, EDSL-web-WASH}.

The semantics of an EDSL is embedded within functions of the host. 
Depending on the implementation two things can happen:
\begin{itemize}
\item In \emph{shallow embedding} the associated semantics is represented directly in the host language.
\item In \emph{deep embedding} a structure representing the domain-specific construct is created, that can later be translated, optimized and run separately.
\end{itemize}

\subsection{Staging}

The type of embedding is closely related to the form of staging supported by the host language.
Staging is a mechanism that controls the execution order of the code:
\begin{itemize}
\item A piece of code may execute within a body of a function that has not been called, often leading to symbolic computation.
\item A piece of code may be kept as a code, despite its surrounding context being executed.
\end{itemize}

The former case is a form of function specialization, the latter -- deferring and code generation.

The simplest approach is \emph{textual staging}, where strings represent fragments of programs.
In \emph{structural staging}, the program code is represented explicitly as a data structure, typically as an Abstract Syntax Tree (AST) or graph.
The structure can be created explicitly, for example through the LLVM instruction builders~\cite{LLVMBuilders}.
Languages such as `C~\cite{TickC} and MetaML~\cite{MetaML} use a dedicated syntax to represent a code object.
Alternatively, the process can be hidden behind overloaded functions, operators, or templates~\cite{CExpr}.

We identify \emph{functional staging} as a case of structural staging where the structure is represented entirely by ordinary functions~\cite{Reynolds75, Tagless}.
In Lightweight Modular Staging~\cite{LMS} these builder functions are hidden by overloading ordinary functions over a higher-kinded type \inline{Rep[T]}.

All the above techniques represent staged code as an object and lead to deep embedding.
MetaML and Light Modular Staging are particularly popular in the context of EDSLs~\cite{LangVirt, Spiral, MetaML-DSL}. 

Not all staging techniques require such a code representation.
Impala~\cite{Impala} and DeepCPS~\cite{DeepCPS} languages achieve the same result by extending the core lambda calculus,
rather than introducing new data types.

With this approach, there is no inherent difference between stages.
For example, DeepCPS can be used to assemble a function incrementally through continuation calls,
and then use staging to transform it to an efficient function as if it was written by hand in one go~\cite{DeepCPS-Actions}.
In this paper we rely on this functional approach to code building.

\subsection{Macro Languages}

The syntactic limitation of embedded DSLs can be lifted, at least partially, through macro processing. 
Simplest macro languages, such as C-preprocessor and m4 operate lexically~\cite{M4}.
Lexical macro languages can be fully programmable, such as TeX~\cite{TeX}.
Advanced macro systems operate with additional syntax knowledge, for example by examining and transforming an AST.
Lisp and Scheme introduced the concept of \emph{hygienic macros}~\cite{Scheme5, LispHygenic},
that are referentially transparent and prevent accidental name capturing.

The macro system of the \inline{<bigwig>} compiler~\cite{GrowMacro} allow grammar extensions to the host language.
The changes can be packaged and templated through so-called \emph{metamorphs}.
However, the macro system is not programmable and recursion is explicitly forbidden.
While macros are hygienic, metamorphs are not.

Macros can be even more powerful when using semantic knowledge.
For example, XL~\cite{Maddox} can be enriched semantically, but its syntax cannot be altered.

\begin{figure*}[tb]  
\begin{tabularx}{\textwidth}{lp{4.4cm}Xr}
& Basic syntax:\\ & $\sfun$ & \inline{(x$_1$, x$_2$, ...)'[y]' \{ $\mathit{b}$\}} & (lambda function) \\
& $\top$ $\bot$ $\AND$ $\OR$ $\NOT$ & \inline{'always'} \inline{'never'} \texttt{\& | !} & (staging constants, operators) \\
& $\Stage{e}v\ \overline{v}$ &
    \inline{'@e:' v v$_1\ $ v$_2\ $ ...} &
    ($\lambda$ body: application) \\
& $\Stage{e}\texttt{fix}\ [y]x = v\ \texttt{in}\ b$ &
    \inline{'@e:' fix '[y]' x} $v$ $b$&
    \text{($\lambda$ body: fix-point combinator)} \\
& Syntactic sugar:\\ & $\lambda\Stage{y}\overline{x} . \Stage{y} v \overline{v}$ &
    \inline{(x$_1$, x$_2$, ...)} \inline{$\ $\{ v v$_1\ $ v$_2\ $ ... \}} &
    (natural staging) \\
& $\Stage{e}\ \left(\lambda\Stage{y}x . b\right)\ v$ &
    \inline{'@e:' let '[y]' $\ x$ $\ v$ $\ b$} &
    (let construct) \\
& $\Stage{e}v\ \overline{v} \left(\lambda[y]x.b\right)$ & \inline{'@e:' v v$_1\ $ v$_2\ $ ... (x$_1$, x$_2$, ...)'[y]' $\mathit{b}$} & (last argument) \\
& $\left(\lambda [y] x . b\right)p \stackrel{\beta}{\longrightarrow} \left(\lambda [y] x . b\right)v$ & \inline{"p" (x)'[y]' $\ b$} & (non-CPS expression $p$) \\
& $\left.v\right|_1, \left.v\right|_2, ... $ & \inline{!v} & (tuple splicing) \\
& $\lambda\Stage{s}\underbrace{\overline{x}}_y . b $ &%
\inline{(!y)'[s]' \{ $\mathit{b}$\}} 
& (tuple aggregate) \\
\end{tabularx}
\vspace{-3mm}
\caption{Comparison between the lambda calculus with Dynamic Staging (left), and the actual syntax of DeepCPS (right).\label{fig:deepcps_syntax}}
\end{figure*}
\subsection{Metamorphic Languages}

Most flexibility is given in what we call a \emph{metamorphic language} ---
a language that allows its user to alter nearly every aspect of the language on the fly, including its syntax and semantics.

This is achieved for example in Racket~\cite{Racket}, a descendant of Scheme.
The programmer can change its syntax, semantic, type system, linking and optimizations.
Racket creates an IR that the user can explicitly alter through \emph{syntax transformers}.
Unfortunately, controlling the transformations is cumbersome.
Several library functions are attributed to do just that\footnote{http://docs.racket-lang.org/reference/stxtrans.html, retrieved on 23.03.2015}.

The macro system of Fortress~\cite{Fortress} allows the user to define nearly arbitrary syntactic constructs,
using the formalism of Parsing Expression Grammars~\cite{PEG}. 
The new constructs can be used within the grammar definition as well, even if it leads to recursion.
SugarJ~\cite{SugarJ} is a language build on top of Java, SDF~\cite{SDF}, and Stratego~\cite{Stratego},
capable of extending Java syntax in a similar way through \emph{sugar libraries}.
The downside of these approaches is that the new constructs must appear at the top level of the source.
They cannot be created dynamically, conditionally, or be parametrized.

To our knowledge, all metamorphic languages operate on a single AST to connect the languages,
either directly or using the deep embedding approaches. 

\section{Overview}
\label{sec:overview}

Our overarching goal is to create a flexible yet comprehensible metamorphic language.
We want syntactic flexibility typical to stand-alone languages.
At the same time, the DSLs should be embeddable in a single host language, permitting communication with other DSLs.
Finally, we expect to generate highly efficient code, without any overhead coming from the language embedding.

Our solution, which we name \emph{ManyDSL}, realizes these goals through the following means:
\begin{itemize}
\item A grammar is used to describe the syntax.
\item The grammar is augmented with parameters passed down and up the productions, as well as semantic action functions put within the rules.
We call this a \emph{syntax-directed execution} scheme, which we describe in \autoref{ch:sde}.
\item The semantic actions are executed immediately during parsing.
We use Dynamic Staging, a feature of DeepCPS, to guide the code generation.
\item Parsing and code execution is interleaved and interconnected, allowing one to affect another.
In particular, new languages can be loaded on the fly and parsing can be switched to those new languages.
\end{itemize}

%
%

\section{The Host Language}

We have chosen DeepCPS as the host language in ManyDSL.
It is a functional language that enforces the Continuation Passing Style (CPS) and enables Dynamic Staging~\cite{DeepCPS}.
\begin{itemize}
\item CPS provides high flexibility when designing custom control flow structures.
All branches and loops can be expressed as functions.
Any DSL embedded in DeepCPS is not limited by the control structures of the host language.
\item Dynamic Staging introduces staging as first-class construct. 
Staging allows the user to specify domain optimization, but also let the new DSLs expose staging in various degrees on their own.
\item DeepCPS allows for incremental building of code, without overhead, through the \emph{fragment functions}~\cite{DeepCPS-Actions}.
\end{itemize}

\subsection{DeepCPS Syntax and Semantics}

In the \autoref{fig:deepcps_syntax} we summarize the syntax of DeepCPS.
Apart from standard lambda calculus semantics with CPS restrictions, DeepCPS adds Dynamic Staging.
It is realized by the \emph{implicit staging parameter} \inline{'[y]'} present in every lambda
and a \emph{staging expression} \inline{'@e:'} present in each lambda body.

When a lambda is invoked, the implicit staging parameter is always replaced by a special staging constant $\top$.
Staging expressions use these staging parameters to form boolean expressions.
When an expression \inline{'@e:'} evaluates to $\top$, the annotated body is considered \emph{active} and is scheduled for execution.
During the execution process, the DeepCPS interpreter maintains a set of all bodies that are active.
At each execution step the deepest active body (containing no nested active bodies) is executed.

Staging variables can be used as normal arguments, and normal variables can appear in staging expressions as well.
Non-stage constants are equivalent to $\top$,
while variables that are still represented only as symbolic values are $\bot$.
The precise formal definition of the syntax and semantics is given in the original DeepCPS paper~\cite{DeepCPS}.

The \inline{!} operator is an extension to DeepCPS that we heavily depend upon.
A parameter preceded by \inline{!} accepts any excessive arguments given to a function, and packs them all into a tuple under the given name.
The symbol \inline{!} in front of an argument unpacks all elements of a tuple and splices them as an argument list to a function call.
For convenience, all arguments that are packed or unpacked through \inline{!}, are highlighted in italics.

\subsection{Staging Chains}
\label{sec:fragment_chaining}

One of the benefits of DeepCPS that we rely on in this paper is the ability to form staging chains.
Each link within the chain is a piece of code that is staged upon some parameter \inline{'@s1:'}.
The last continuation introduces a new implicit staging parameter \inline{'[s2]'} that is used to stage another link
\begin{code}[b]
\begin{lstlisting}
  '@s1:' fct !args (!params)'[s2]'
	...
	'@s2:' fct2 !args2 (!params)'[s3]'
	...
	'@s3:' fct3 !args3 (!params)'[s4]'
	...
\end{lstlisting}
\caption{\label{code:staging_chain}
A series of functions calls chained together by a series of staging parameters \inline{'s1'}, \inline{'s2'}, \inline{'s3'}, \inline{'s4'}.
Such a staging chain \inline{'s'} is executed by triggering the \inline{'s1'} variable.
Afterward, all pieces are executed in order, sequentially activating the next staging parameter.}
\end{code}
A series of links of that kind put together, as for example in \autoref{code:staging_chain}, form a chain $s$.
After the first link is executed the following links become invoked as well, in sequence.

The links in the example follow one another.
However, because the staging variables can be freely passed between functions, the links can originate from completely independent functions.

\subsection{Building Code}
\label{sec:building_code}

Staging can be used to build arbitrary function incrementally.
Each incremental addition is kept in a separate lambda of the form:
\begin{lstlisting}
(!args, cont) ... "computation" ... cont !args2
\end{lstlisting}
The \inline{!args} is a set of arbitrary arguments passed between the incremental additions.
The \inline{cont} is the continuation representing the rest of the function we construct.
We refer to these lambdas as \emph{subject code} as they contain the code of a function we build.
These lambdas are connected together through the builder functions:
\begin{itemize}
\item The \inline{build} function takes the subject code as a lambda argument and encapsulates it in a \emph{fragment function}.
\item The \inline{merge} function takes two fragment functions and connects them together.
The end result is a new, bigger fragment function containing the subject code of both arguments merged together.
\end{itemize}

Usually, each code needs a single continuation.
However, when the subject code represents a branch or a jump, the number of continuations may differ.
We refer to that number as \emph{arity} of a fragment function.
The arity must be provided by the user into the \inline{build} function.

\begin{code}
\begin{tikzpicture}[xscale=0.169,yscale=-0.333]
\highlightcode{blue}{4}{2}{50}{8}{};
\highlightsolid{white}{6}{4}{50}{5};
\highlightsolid{white}{6}{6}{50}{7};
\highlightcode{red}{4}{10}{50}{16}{};
\highlightsolid{white}{6}{12}{50}{13};
\highlightsolid{white}{6}{14}{50}{15};
\highlightcode{darkgreen}{29}{17}{42}{18}{};
\highlightcode{darkgrey}{29}{18}{42}{19}{};
\highlightcode{darkbronze}{29}{19}{43}{20}{};
\node[inner sep=0,anchor=north west] at (0,-0.2) {
\begin{lstlisting}
let create_signum(return) {
	build 2 ('ft', val, exit, cont1, cont2)'[bt]' {
		'@ft:' "val>0" (positive)
		if positive ()'[ft]' {
			'@bt:' cont1 'ft' val exit
		} ()'[ft]' {
			'@bt:' cont2 'ft' val exit
		}
	} (Fif_pos)
	build 2 ('ft', val, exit, cont1, cont2)'[bt]' {
		'@ft:' "val<0" (negative)
		if negative ()'[ft]' {
			'@bt:' cont1 'ft' val exit
		} ()'[ft]' {
			'@bt:' cont2 'ft' val exit
		}
	} (Fif_neg)
	build 0 ('ft',val,exit)'[bt]' { '@ft:' exit 1 } (Fp)
	build 0 ('ft',val,exit)'[bt]' { '@ft:' exit 0 } (Fz)
	build 0 ('ft',val,exit)'[bt]' { '@ft:' exit -1 } (Fn)
	merge Fif_pos Fp (F)
	merge F Fif_neg (F)
	merge F Fn (F)
	merge F Fz (F)
	finalize F P
	return (arg, exit)'[ft]' {
	  '@P:' P 'ft' arg exit
	}
}
\end{lstlisting}
};
\end{tikzpicture}
\caption{\label{code:building_code}
Example of building a code for a function \inline{signum}.
}
\end{code}

In order to see how the builders work in practice, consider an example in \autoref{code:building_code}.
Here a function \inline{signum} is being constructed, defined as:
$$\mathrm{signum}(x) = \left\{
\begin{array}{rl}
1&x>0\\
0&x=0\\
-1&x<0\\
\end{array}
\right.$$
The body of the function consists of two conditionals, checking if the argument is positive, negative or neither.
Depending on a result, a different value is returned through the \inline{exit} continuation.

All the function-related computation is performed in \inline{'ft'} staging chain.
The staging chain \inline{'bt'} is responsible for calling all the merged continuations early, so that they are removed from the produced code.
In the end, we obtain a function given in \autoref{code:signum}.
\begin{code}
\begin{tikzpicture}[xscale=0.169,yscale=-0.333]
\highlightcode{blue}{2}{1}{50}{11}{};
\highlightcode{red}{4}{5}{50}{10}{};
\highlightcode{darkgreen}{4}{3}{34}{4}{};
\highlightcode{darkbronze}{6}{7}{34}{8}{};
\highlightcode{darkgrey}{6}{9}{34}{10}{};

\node[inner sep=0,anchor=north west] at (0,-0.2) {
\begin{lstlisting}
(arg, exit)'[ft]' {
  '@ft:' "val>0" (positive)
	if positive ()'[ft]' {
		'@ft:' exit 1
	} ()'[ft]' {
	  '@ft:' "val<0" (negative)
	  if negative ()'[ft]' {
	    '@ft:' exit -1
	  } ()'[ft]'
	    '@ft:' exit 0
	}
}
\end{lstlisting}
};
\end{tikzpicture}
\caption{\label{code:signum}
The code produced by the builders from \autoref{code:building_code}.
Only code in the \inline{'ft'} staging chain remains.
We get a tight representation and
no overhead from the construction process.
}
\end{code}

The precise definition of \inline{build} and \inline{merge}, together with the explanation how they work is given in~\cite{DeepCPS-Actions}.

\section{Functional Grammar}

When discussing functional parsing, one typically thinks about parser combinators (\autoref{ch:parcomb}).
The generated parser is treated as an ordinary function.
It is up to the parser creator to ensure that it is efficient, but this is non-trivial.
The programmer must understand when exactly lazy evaluation is triggered and how combinators need to be connected to avoid ambiguities.

In our approach, we use a functional language to create a \emph{grammar} instead of a parser.
Once created, the grammar is then processed in a traditional way to create an LL(1) parser.
Although it may seem as a step backwards, this allows us to maintain important practical properties:
\begin{itemize}
\item Backtracking is guaranteed to never to occur.
\item Any ambiguities are detected when the parser is generated.
\item The parsing process is straightforward and easy to follow.
\end{itemize}
On the other hand, since the grammar is defined within a functional language, we still maintain composability similar to when using parser combinators.

\subsection{Syntax-Directed Execution}
\label{ch:sde}

We use a new \emph{syntax-directed execution} scheme (SDE) as a basis of our parsing process.
On the surface, it is very similar to syntax-directed translation scheme (SDT)~\cite{SyntaxTransduction,GrammarParadigms}.
SDE however puts emphasis on the execution of code. 
There are no objects representing the parse tree, AST, or IR that would be generated.
Instead, productions and actions are treated as functions that are executed as a part of the parsing process.
Code can be generated during parsing through lambda encapsulation and staging.

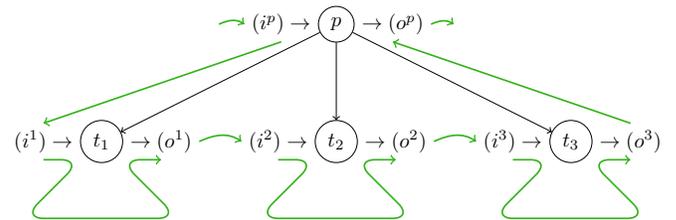
\begin{figure}[b]
\scalebox{0.78}{%
\begin{tikzpicture}
\draw (0,4) node(node)[circle,draw] {$p$};
\node(in)[left] at (node.west) {$(i^p) \rightarrow$};
\node(on)[right] at (node.east) {$\rightarrow (o^p)$};

\draw (-4,2) node(t1)[circle,draw] {$t_1$};
\node(i1)[left] at (t1.west) {$(i^1) \rightarrow$};
\node(o1)[right] at (t1.east) {$\rightarrow (o^1)$};

\draw (0,2) node(t2)[circle,draw] {$t_2$};
\node(i2)[left] at (t2.west) {$(i^2) \rightarrow$};
\node(o2)[right] at (t2.east) {$\rightarrow (o^2)$};

\draw (+4,2) node(t3)[circle,draw] {$t_3$};
\node(i3)[left] at (t3.west) {$(i^3) \rightarrow$};
\node(o3)[right] at (t3.east) {$\rightarrow (o^3)$};

\draw [->] (node) to (t1);
\draw [->] (node) to (t2);
\draw [->] (node) to (t3);

\path [darkgreen,->,thick,out=25,in=145] (-2,4) edge (in.west);
\path [darkgreen,->,thick] (in.south) edge (i1.north);

\draw [darkgreen,->,thick,rounded corners=10pt] (i1.south) -- ++(0.65,0) -- ++(-1,-1) -- ++(+2.65,0) -- ++(-1,+1) -- (o1.south);
\path [darkgreen,->,thick,out=25,in=145] (o1.east) edge (i2.west);
\draw [darkgreen,->,thick,rounded corners=10pt] (i2.south) -- ++(0.65,0) -- ++(-1,-1) -- ++(+2.65,0) -- ++(-1,+1) -- (o2.south);
\path [darkgreen,->,thick,out=25,in=145] (o2.east) edge (i3.west);
\draw [darkgreen,->,thick,rounded corners=10pt] (i3.south) -- ++(0.65,0) -- ++(-1,-1) -- ++(+2.65,0) -- ++(-1,+1) -- (o3.south);

\path [darkgreen,->,thick] (o3.north) edge (on.south);
\path [darkgreen,->,thick,out=25,in=145] (on.east) edge (+2,4);

\end{tikzpicture}
}
\caption{A fragment of attributed Abstract Syntax Tree created for a production $p \rightarrow t_1 t_2 t_3$.
An L-attributed tree can be traversed in a depth-first left-to-right fashion (green path).
Each attribute along the path depends only on the values encountered earlier on the path.
\label{fig:L-attrib-AST}
}
\end{figure}

The language is defined by an L-attributed~\cite{LAttribute} LL(1) grammar augmented by semantic actions that may appear at any position in the production body.
A hypothetical parse tree can be evaluated with a single depth-first left-to-right traversal, as shown in \autoref{fig:L-attrib-AST}.
In our SDE scheme, however, no tree is ever generated. Grammar terms are treated as function calls.
Attributes are replaced with an equivalent notion of parameters and arguments that are passed into and from the terms.
We represent each attributed term $t$ as $(i)\rightarrow t \rightarrow (o)$ to indicate the flow of the data --- input arguments (i) are passed into $t$ and the results are returned into $(o)$.  A complete parametrized production of the form $p ::= t_1 t_2 t_3$ looks as:
\begin{align*}
{\color{red}(i^p)}\rightarrow p \rightarrow {\color{darkgreen}(o^p)} ::=\ & {\color{darkgreen}(i^1)}\rightarrow t_1 \rightarrow {\color{red}(o^1)}\\
& {\color{darkgreen}(i^2)}\rightarrow t_2 \rightarrow {\color{red}(o^2)}\\
& {\color{darkgreen}(i^3)}\rightarrow t_3 \rightarrow {\color{red}(o^3)}
\end{align*}

The head of the production acts as a function header (or signature):
The input values are parameters ({\color{red}red}) --- a list of names, that are set to concrete values when the production is taken.
The output values are return arguments ({\color{darkgreen}green}) that must be concrete values themselves by the time the production is resolved.
This distinction is reversed for each term within the body: input values are \emph{arguments} and output entries are \emph{parameters} which become set by the called production.
A name at argument position always refers to the same name last seen at the parameter position.

A grammar may feature many productions for the same nonterminal.
However, the signature of all productions for given nonterminal must be equivalent:
the name and number of the input parameters must be the same, as well as the number of output values.

The set of all productions for a given nonterminal $p$ is refered to as a \emph{rule} of $p$.
When such a rule is invoked through its head nonterminal, the parser performs a standard LL1 lookahead to decide which particular production should be taken.

\subsection{Language Programming Interface}

Language grammar defines one or more \emph{entry rules}.
These are the possible rules where the parsing in given language begins.

Entry rules may be referred within productions of other languages.
For example, if in language $A$ a language $B$ is used through an entry rule of nonterminal $n$, we denote it as:
$${\color{darkgreen}(i^1)}\rightarrow B.n \rightarrow {\color{red}(o^1)}$$
When such a \emph{foreign nonterminal} is encountered, the parser completely switches the language:
The rules of $A$ become nonexistent and only rules of grammar $B$ are in effect.
When the entry rule of $B$ finishes, the parser switches back to language $A$.

In general, foreign nonterminal may appear anywhere a normal nonterminal would.
The only restriction is that a token that is a foreign nonterminal cannot be used to compute the parse table for LL1.
Grammar $A$ has no knowledge of tokens of grammar $B$. These two grammars may use a completely different set of tokens.

The parser does not create any global AST.
The only information that is exchanged between the languages is defined by the foreign nonterminal call and the respective entry rule.
For that reason, we refer to the set of all entry rules for given language as a \emph{Language Programming Interface} (LPI).
Knowing the LPI suffices to use the given language with others.
Other language rules and actions are private to that language.

\subsection{Abstractions over Grammar}

\begin{code}
\begin{lstlisting}
function lassoc<elem, op, action> {
	N ::= elem R;
	R ::= epsilon;
	R ::= op elem action R;
	return N;
}
function rassoc<elem, op, action> {
	N ::= elem R;
	R ::= epsilon;
	R ::= op elem R action;
	return N;
}
grammar MinusDiv {
  Diff ::= lassoc< Quotient, ""-"", epsilon >;
  Quotient ::= lassoc< Value, ""/"", epsilon >;
}
\end{lstlisting}
\vspace{-2mm}
\caption{\label{code:associativity}Grammar functions for a left and right associative binary operators. The difference is in the position of the action use.
We use \inline{lassoc} to create a grammar supporting left-associative \inline{-} and \inline{/} operators and taking precedence into the account.}
\end{code}
\begin{code*}
\begin{subfigure}[b]{0.49\textwidth}
\begin{lstlisting}
function lassoc<elem, op, action> {
  alias |v| = |elem:out|;
  N|->(v)| ::= elem|->(v)| |(v)->|R|->(v)|;
	|(v)->|R|->(v)| ::= epsilon;
	|(v)->|R|->(v)| ::= op
	                elem|->(r.v)|
		              |(v,r.v)->|action|->(v)|
		              |(v)->|R|->(v)|;
	return N;
}
\end{lstlisting}
\end{subfigure}
\begin{subfigure}[b]{0.49\textwidth}
\begin{lstlisting}
|(S)->|Value|->(S)|
lassoc<Value, ""/"", action> $\text{\textrm{reduces to:}}$
§(S)->§N|->(S)| ::= §(S)->§Value|->(S)| |(S)->|R|->(S)|;
|(S)->|R|->(S)| ::= epsilon;
|(S)->|R|->(S)| ::= op
                §(S)->§Value|->(r_S)|
	              |(S,r_S)->|action|->(S)|
	              |(S)->|R|->(S)|;
\end{lstlisting}
\end{subfigure}
\caption{\label{code:lassoc-args}
The \inline{lassoc} abstraction taking the rule arguments into the account.
On the right, an example derivation of the abstraction for the element rule \inline{|(S)->|Value|->(S)|}.
Almost all term arguments are obtained through simple substitution, but the input arguments for \inline{Value} (marked in {\color{langdslargdefault}pink})
are added later as the default arguments. The input parameter for \inline{N} is added too, because otherwise the first \inline{S} would be undefined in the production. Note that two stacks are passed into \inline{action}, while only the second one should actually be used.
}
\end{code*}

We created a library and a language on top of DeepCPS that facilitates the creation of new grammars.
Our new language, LangDSL, uses a syntax similar to one used to describe the SDE in \autoref{ch:sde}.
The productions do not have to appear at any specific point in the code.
While grammar, productions, and terms carry additional semantic value, they can be passed freely between functions.
This allows us to define abstractions over portions of the grammar.
The grammar abstraction \inline{f} can be defined through \inline{function f<params>} syntax and later used within the context of another production through the invocation \inline{f<args>}.

A typical problem for LL parser is the handling of left recursion.
This is most common when parsing an expression that uses left-associative binary operators.
A well known solution is to take the offending production:
$$A ::= A\alpha | \beta$$
and rewrite it as:
\begin{align*}
A & ::= \beta R\\
R & ::= \epsilon | \alpha R
\end{align*}
The new grammar can parse the same input, but creates a different AST that resembles the use of a right-associative operator.

In SDE however we do not create an AST thus the above is not a problem.
The distinction between left and right association is achieved not through the shape of the tree,
but through the position of the semantic action.
Depending on its position, the actions are executed in different order, taking different arguments.

Consider an example in the \autoref{code:associativity} were we create two grammar abstractions --- one for each: left and right associative binary operator.
In the left-associative operator, the action is performed before the recursive call. As a result, actions are performed in the same order as input is being read.
In the right-associative operator, action is performed after the recursion is complete.
They are performed in the order of productions returning, starting from the bottom.

We use the abstractions to create concrete grammar for binary \inline{-} and \inline{/}.
In our example, the grammar recognizes any expression using the two operators, but performs no semantic action yet.


\subsection{Abstractions over Parameters}

Grammar abstractions are higher-order functions. Each argument, such as \inline{elem} in the example above is a name of a nonterminal,
which in turn acts as another function.
The \inline{elem} argument may take some arguments, perform parsing operation, and return a new set of values.
Notice however, that in the abstraction in \autoref{code:associativity} we did not specify what arguments \inline{elem} may have.

This is not an accedient. We would like the functions such as \inline{lassoc} to be generic enough so that all versions of \inline{elem} are accepted. Consider, for example different flavors of \inline{MinusDiv} grammar having values of different kind:
\begin{itemize}
\item A value may simply be a single integer returned by the \inline{Value} rule: \inline{|->(integer)|}
\item A value may be a quotient represented by a pair of integers. A \inline{Value} would return a pair: \inline{|->(num, denom)|}.
\item A value may be a string name. In order to retrieve a value an environment parameter \inline{env} may be needed. Such use example is discussed in \autoref{ch:env}.
\item Elements may be added to a stack \inline{S}. We would then have a production of the form \inline{|(S)->|Value|->(S')|}.
\end{itemize}

LangDSL helps achieve that goal in three ways:

\begin{itemize}
\item First, when an argument or parameter is missing, LangDSL adds a \emph{default} argument -- it is the name of the corresponding parameter causing the error.
This is performed after all grammar abstractions are resolved.
Only the entry rules, that define the LPI, cannot be altered in that way.
\item Secondly, a grammar abstraction can take an argument defining the name list.
Whenever the name list is used as a nonterminal's input or output, the content of the list is spliced.
Moreover, for a name tuple \inline{n}, a value \inline{prefix.v} is a new name tuple with a \inline{prefix_} added to each element of the original \inline{n}.
\item Finally, for any grammar term \inline{t}, its input and output argument list can be checked throuh \inline{|t:in|} and \inline{|t:out|}.
\end{itemize}

While such string manipulation is a simple solution to the problem, in most cases it suffices.
Any potential name clashes can be avoided by adding the prefix, and the scope of the names is local.
With this help \inline{lassoc} can be made generic enough, as shown in \autoref{code:lassoc-args}, to handle all the cases given above.

\section{Semantic Actions}
\label{ch:actions}

Semantic actions are small DeepCPS functions embedded within the grammar.
These actions are invoked whenever the parser reaches the term within the production.
Upon invocation the parsing process is halted and the action function is executed.

As shown in \autoref{code:action}, an action has an arbitrary number ($n$, $m$) of input and output attributes.
The DeepCPS function must take $n+1$ parameters.
It also provides an implicit staging parameter \inline{'parse'} that is set to $\top$ when the parser invokes the function.
The action's input arguments are mapped to the first $n$ parameters.
The $n+1$ argument is a \inline{return} continuation function provided by the parser.
Calling the continuation returns the control back to the parser.
The \inline{return} accepts $m$ arguments that are passed back into the output parameters of the action.

\begin{code}
\begin{lstlisting}
$\color{langdslarg}(i_1, i_2, ..., i_n)$ |->| $\color{langdslarg}(o_1, o_2, ..., o_m) $ ▼{
  ▼'@parse:'▼ ... $\color{langdsldeepcps}\mathrm{body}$ ...
	return $\color{langdsldeepcps}o_1\ o_2\ ...\ o_m$
}▼\end{lstlisting}
\vspace{-2mm}
\caption{A semantic action with $n$ input and $m$ output attributes, which are mapped to DeepCPS function and its continuation.
For convenience, the action body is put after, and not in between input and output parameters.
The DeepCPS section between the curly braces ({\color{langdsldeepcps}green}) skips the head of a lambda, as it is auto-generated as:
\inline{(i$_1$, i$_2$, ..., i$_n$, return)'[parse]'}.
\label{code:action}
}
\end{code}

\subsection{Immediate Execution}

The most straightforward use of the semantic actions is to provide the intended semantic meaning of the parsed code directly.
The \inline{MinusDiv} grammar with all the necessary parameters and actions is given in \autoref{code:immediate}.
The action takes two arguments $l$ and $r$, and applies the corresponding mathematical operation at that point in time, during parsing.
The result is returned as $v$ back to the parser and used in subsequent production calls.

\begin{code}[t]
\begin{lstlisting}
grammar MinusDiv {
  Diff|->(v)| ::= lassoc< Quotient, ""+"",
            |(l,r)->(v)| ▼{
              ▼"l-r"▼ (diff)
		  				return diff
            }▼ >;
  Quotient|->(v)| ::= lassoc< Integer, ""*"",
            |(l,r)->(v)| ▼{
              ▼"l/r"▼ (quot)
		  				return quot
            }▼ >;
}
\end{lstlisting}
\vspace{-2mm}
\caption{Grammar for binary \inline{-} and \inline{/}, performing the computation immediately, during parsing.
The final result is a single number.
\label{code:immediate}}
\end{code}

\subsection{Code Generation}

The SDE scheme can also be used for code generation, through means of Dynamic Staging and builders given in \autoref{sec:building_code}.
In \autoref{code:builded} there is no change in the grammar itself, but the parameters and actions are a bit different.

\begin{code}[t]
\begin{lstlisting}
▼let finalize(F,return)▼'[bt]'▼ { return
  (!args)▼'[ft]'▼
	  ▼'@bt:'▼ F ▼'always'▼ ▼'ft'▼ !args
}▼
grammar MinusDiv {
  Expr|->(P)| ::=
    |()->(F)| ▼{
	    build 1 (!args, cont){cont !args} return
  	}▼
  	|(F)->|Diff|->(F)|
  	|(F)->(P)| ▼{
  	  build 0 (▼'ft'▼, v, end) {▼'@ft:'▼ end v} (Fend)
  		merge F Fend (F)
  		finalize F return
  	}▼;
	
  |(F)->|Diff|->(F)| ::= lassoc< Quotient, ""+"",
          |(F)->(F)| ▼{
	    		  build 1 (▼'ft'▼, l, r, !args, cont)▼'[bt]'▼ {
              ▼'@ft:'▼ ▼"l-r"▼ (diff)▼'[ft]'▼
		    		  ▼'@bt:'▼ cont ▼'ft'▼ diff !args
		  	  	} (Fnext)
		  		merge F Fnext return
          }▼ >;
  |(F)->|Quotient|->(F)| ::= lassoc< Number, ""/"",
          |(F)->(F)| ▼{
	  				build 1 (▼'ft'▼, l, r, !args, cont)▼'[bt]'▼ {
              ▼'@ft:'▼ ▼"l/r"▼ (quot)▼'[ft]'▼
		  				▼'@bt:'▼ cont ▼'ft'▼ quot !args
		  			} (Fnext)
		  			merge F Fnext return
          }▼ >;				
  |(F)->|Number|->(F)| ::= Integer|->(v)|
          |(F,v)->(F)| ▼{
	  				build 1 (▼'ft'▼ !args, cont) {
	  					cont ▼'ft'▼ v !args
	  				} (Fnext)
	  				merge F Fnext return
	  			}▼;
}
\end{lstlisting}
\vspace{-2mm}
\caption{Grammar for binary \inline{-} and \inline{/} with a deferred computation.
The result is a function that, when invoked, performs the computation.
\label{code:builded}}
\end{code}


The bodies of the actions still contain the same code: a mathematical operation followed by a continuation call.
This time however, these operations are embedded as a subject code within the fragment functions.
 
Note how number literals are handled in the \inline{Number} production.
A single line within the builder takes the new value \inline{v} and concatenates it into the recurring \inline{!args} tuple
by calling the continuation \inline{cont} with arguments \inline{ft v !args}.
As a result, each time a number is read, the arity of \inline{!args} increases by one.
The \inline{-} and \inline{/} operations pop two last elements of \inline{!args} and push the result, decreasing the arity of the tuple by one.

Note that all these operations on \inline{!args} are \emph{not} staged in the \inline{'ft'} chain.
They are resolved early, during the construction, substituting the respective arguments of the \inline{-} and \inline{/} operators.
As a result, the final code contains only the mathematical operators, without any overhead.

The new production \inline{Expr} initializes a new function fragment representing the expression.
At the end, it finishes it with the call to \inline{finalize}, which triggers the \inline{'bt'} chain (set to \inline{'always'}), but leaves the \inline{'ft'} chain intact. It returns a lambda \inline{(!args)'[ft]'} with only function-time code in it.
For the input \inline{1-4/2-3} we obtain:
\begin{lstlisting}
(end)'[ft]' {
  '@ft:' "4/2" (quot)'[ft2]'
	'@ft2:' "1-quot" (diff1)'[ft3]'
	'@ft3:' "diff1-3" (diff2)'[ft4]'
	'@ft4:' end diff2
}
\end{lstlisting}

\subsection{Multi-domain code}

Fragment functions, same as any other data, may be passed through the LPI into another language.
For each fragment $F$ the languages must agree on the type of fragments accepted. This entails:
\begin{itemize}
\item the minimum expected arity of $F$
\item the arguments that are passed into the subject code of $F$ through the \inline{!args}.
\end{itemize}

Note that the languages do not need to agree on the inner representation of the code.
The grammar of one language, as well as the structure of the fragment functions that were used to construct $F$ does not matter for the other.

For that reason, even in the context of fragment functions, it suffices to check the signatures of the entry rule and rely on the LPI.
The designers of the languages may choose to exchange more complex data structures,
such as a representation of compound or recursive data types.
This is however a decision to be made by the language designers, independently from ManyDSL core.
It is not forced by ManyDSL itself.

\section{Challenges}

In the previous section we have explained the basic mechanism of the Syntax-Directed Execution scheme with the use of Dynamic Staging.
Let us now focus on more pragmatic challenges when designing a small DSL.

\subsection{Custom Environment}
\label{ch:env}

One of important aspects of almost any language is name binding.
What scopes does a DSL provide and how can names be mapped to values?
How is the name lookup performed between languages?

With our SDE scheme we do not have to rely on a single approach.
Name binding can be handled by user code, possibly in an early stage to avoid run-time overhead.
Let us assume that the environment is represented in an object-oriented style, through a mutable object \inline{env}
with methods \inline{insert} and \inline{lookup}.


\begin{code}[b]
\begin{lstlisting}
|(F)->|Assgn|->(F)| ::= Identifier|->(id)| ""=""
        Expr|->(P)| "";""
				|(id,F,P)->(F)| ▼{
					build 1 (▼'ft'▼, !args, env, cont)▼'[bt]'▼ {
					  ▼'@ft:'▼ P env (v)▼'[ft]'▼
						▼'@bt:'▼ ▼"env.insert(id,v)"▼ (env)
						cont ▼'ft'▼ !args env
					} (Fnext)
				  merge F Fnext return
				}▼;
|(F)->|Number|->(F)| ::= Identifier|->(id)|
        |(F,id)->(F)| ▼{
					build 1 (▼'ft'▼, !args, env, cont) {
					  ▼"env.lookup(id)"▼ (v)
						cont ▼'ft'▼ v !args env
					} (Fnext)
					merge F Fnext return
				}▼;
\end{lstlisting}
\vspace{-2mm}
\caption{Example use of an environment within builders.
This grammar extends \autoref{code:builded} to support named values.
In the produced code name binding is already resolved as no environment operations are present in the staging chain \inline{'ft'}.
\label{code:env_use}}
\end{code}

Custom name binding can be realized by such environment, within the builders, staged in the build-time chain.
By doing so, function-time values are stored symbolically within the environment and are referenced as such in other fragments.
For example, in \autoref{code:env_use} we use the environment to extend the \inline{MinusDiv} grammar to support assignment statement \inline{id = Expr}.
The \inline{Expr} rule builds a function \inline{P} that is used to compute a value \inline{v} at function-time.
However, at build time we take the symbolic name \inline{v} and include it in the environment under a new name \inline{id}.

The identifiers can be used as a part of an \inline{Expr}, replacing the constant integers.
When an identifier is encountered, still at build time, the symbolic name \inline{v} is retrieved.

In the above example we use a single environment type within a single DSL.
However, the implementation of each environment is independent and may be very different from another one.
As long as they share the signature, they can be combined together.

Consider an \inline{env} object with a dynamic dispatch for its methods that implement language $A$ lookup rules.
It is then passed through the LPI to another language $B$.
Then, if the language $B$ uses the object, it can access values defined in language $A$ using $A$'s scoping rules provided by the polymorphic object \inline{env}.

\subsection{Multi-pass Language}
\label{ch:circular}

Some DSLs require multiple passes over their AST for semantic analysis and code generation.
This is most common when a DSL supports some form of recursion where all declared terms must be visible before processing their definitions.
In our SDE scheme we do not create an AST that could be traversed.
However, each action can create multiple fragments that can be connected in some different order.

Consider a simple example DSL for specifying directed graphs.
Each entry consist of a head vertex, and an edge list naming the vertices where the head connects to.
We want to represent the graph as an adjacency tuple, with each vertex being implicitly represented as an index.
With the input of the form
\begin{lstlisting}
Start -> X, Y;
X -> Y;
Y -> X, Start;
\end{lstlisting}
we want to create an environment:
\begin{lstlisting}
[[""Start"",1],[""X"",2],[""Y"",3]]
\end{lstlisting}
and an indexed list describing the graph, such as:
\begin{lstlisting}
[[2,3],[3],[2,1]]
\end{lstlisting}

The full grammar is given in \autoref{code:graph-full} in the Appendix.
Let us consider here only actions that must be performed when reading the head vertex and when reading the edge list.
We create two fragments: \inline{Decl} and \inline{Def}. Upon reading the head, we change our environment by assigning a new index to the vertex name:
\begin{lstlisting}
|(Decl,name)->(Decl)| ▼{
  build 1 (▼'ft'▼, env, idx, end, cont) {
    ▼"env.insert(name,idx)"▼ (env)
		▼"idx+1"▼ (idx)
		cont ▼'ft'▼ env idx end
	} (DeclNext)
	merge Decl DeclNext return
}▼
\end{lstlisting}
When reading a name within the edge list we look-up the index within the environment and update the adjacency list:
\begin{lstlisting}
|(Def,name)->(Def)| ▼{
  build 1 (▼'ft'▼, env, graph, adjacent, end, cont) {
    ▼"env.lookup(name)"▼ (idx)▼'[bt]'▼
    ▼'@ft:'▼ ▼"concat(adjacent,[idx])"▼ (adjacent)▼'[ft]'▼
    ▼'@bt:'▼ cont ▼'ft'▼ env graph adjacent end
  } (DefNext)
  merge Def DefNext return
};
\end{lstlisting}
Note that \inline{Decl} fragments are merged to other \inline{Decl}-s, and \inline{Def} only to other \inline{Def}-s.
Only at the end, when the whole graph has been parsed, the \inline{Decl} and \inline{Def} fragments are connected.
This way we obtain a single function where all declarations appear before the adjacent list definitions.

\subsection{Type System}
\label{ch:challenge:type}

The type system of DeepCPS is rudimentary and provides no static correctness with the respect to staging.
Using it directly in a higher-level DSL would be limiting, and the produced error messages originating from within the semantic actions could be confusing to the DSL user.

However, type system need not be limited to a fixed rule set of a language.
Type interference can be seen as a form of partial evaluation of the code with respect to its type annotations~\cite{AuxiliaryComp}.
In a DSL embedded in DeepCPS we represent this as auxiliary values (e.g. types) and auxiliary computation performed in an early stage.

For a simple example, consider an extension to the \inline{MinusDiv} language supporting different types of numbers.
Each DSL value is represented by two variables: the actual value and its type. 
Before each mathematical operation we first check the operand types and compute the type of the result.
For example, within the action of \inline{Diff} production of \autoref{code:builded} we have:
\begin{lstlisting}
build 1 ('ft',lval,ltype,rval,rtype,!args,cont)'[bt]' {
	'@ltype & rtype:'            //$\text{\textrm{as soon as types are known}}$
	"ltype != rtype" (error)
	if error () {                       //$\text{\textrm{error occurred}}$
	  print ""Type mismatch!"" ()
		exit
	} ()                            //$\text{\textrm{else --- no type error}}$
	let '[ok]' difftype ltype
  '@ok & ft:' "lval-rval" (diff)'[ft]'
  '@bt:' cont ft diff difftype !args
} (Fnext) ...
\end{lstlisting}

This way type checking is performed in the code itself.
In this simple example it boils down to a simple comparison, but a custom DSL may perform more involved checks.

Note that we chose not to create a dedicated stage chain for type checking (e.g. \inline{'tc'}).
Instead, the check is performed as soon as all the necessary information is available.
This way, the same function can be used in a statically-typed and dynamically-typed DSL,
as well as in a context of type-dependent functions.

\section{Implementation}

\subsection{Structure}

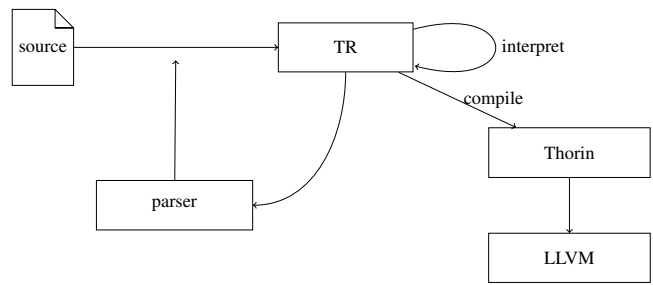
\begin{figure}[b]
\scalebox{0.70}{
\begin{tikzpicture}
\tikzstyle{code}=[shape=rectangle,draw,inner xsep=30pt,inner ysep=10pt]
\draw (0,4) node(source)[document] {source};
\draw (5.75,4) node(TR)[code] {TR};
\draw (2.5,1) node(parser)[code] {parser};
\draw (10,2) node(Impala)[code] {Thorin};
\draw (10,0) node(LLVM)[code] {LLVM};
\draw [->] (source) edge node(mid)[below]{} (TR);
\draw [->] (TR) edge[loop right] node{interpret} (TR);
\draw [->] (TR) edge node[right]{compile} (Impala);
\draw [->] (Impala) edge (LLVM);
\draw [->] (TR) edge [out=270,in=0] (parser);
\draw [->] (parser) edge (mid);
\end{tikzpicture}
}
\caption{The structure of ManyDSL. Source code is parsed directly into Target Representation (TR).
TR can be interpreted and partially evaluated, producing a more efficient TR code.
It can also modify the parser of ManyDSL, so that different DSLs can be read.
Finally, TR code can be translated to Thorin~\cite{Thorin} and then compiled to LLVM~\cite{LLVM}.
\label{fig:ManyDSL-structure}}
\end{figure}

ManyDSL is the name we have given to our implementation of the metamorphic language described in this paper.
The ManyDSL workflow is sketched in \autoref{fig:ManyDSL-structure}.
The first step is parsing the source code into Target Representation (TR).
TR is interpreted and partially evaluated using Dynamic Staging.
By calling a built-in \inline{▲compile}, any function can be compiled, through Thorin~\cite{Thorin} and LLVM~\cite{LLVM}, into highly efficient machine code.

Initially, ManyDSL can parse only DeepCPS source, which has almost 1:1 correspondence to TR.
The user however can introduce a new language using the SDE scheme presented in this paper.
When the language definition is executed by ManyDSL and LL parser is created, it can be used to parse the remainder of the source code.

While technically possible, the user never directly creates nor modifies TR.
It is transformed solely through the semantics of lambda calculus with dynamic staging.
The partially evaluated TR code can also be emitted back as DeepCPS.
We use this mechanism for bootstrapping LangDSL (\autoref{ch:langdsl}), but it is not needed in the normal usage of ManyDSL.

\subsection{Interleaved Execution}

It would be pointless to change the parser when all source is already read.
For that reason, in ManyDSL the parsing process can be interrupted letting the ManyDSL interpreter execute the part that has been already translated into TR.
This interleaved parsing and execution gives ManyDSL a unique possibility where the code and the parser can communicate with each other.

This communication is primarily used for user-guided parsing, but it can benefit other pragmatic situations.
For example, the name of an include file may be the result of some computation --- such as inspecting the operating system environment or the target hardware architecture.
This way DeepCPS code may contain not only the program to be compiled, but the whole build system around it.

In DeepCPS the interruption is achieved through a special syntax \inline{##}.
However, as explained in \autoref{ch:actions}, semantic actions are executed immediately during parsing.
That means, switching between parsing and execution occurs implicitly at every action.

\subsection{Language Library}
\label{ch:langdsl}

The parser generator of ManyDSL is available through C API.
The grammar is build incrementally by building its structure, piece-by-piece by invoking these C functions from DeepCPS.
The C functions can take higher-order DeepCPS values as arguments.

We used this low-level approach to bootstrap LangDSL with the syntax explained in this paper.
With LangDSL, the use of C API is entirely hidden from the user.
Circular references between nonterminals are handled the same way as shown in \autoref{ch:circular}.
Since the bodies of semantic actions are written in DeepCPS, we instruct ManyDSL to switch parsers from LangDSL to native DeepCPS whenever an action is encountered.

LangDSL includes a few additional syntactic sugar constructs that were omitted in this paper to further simplify language creation.
It includes a dedicated syntax for the build+merge pattern.
Commonly recurring production parameters, such as $F$ in \autoref{code:builded}, can be skipped at the use site of a nonterminal.

\subsection{Language Switching}

The output of LangDSL is a function describing a new language.
When invoked, the parser generator of ManyDSL is used to create a new LL(1) parser.
Afterward, when the user orders ManyDSL to change the parser, the next token in the input stream is processed by the new language.

At low level, switching is performed by calling C functions from DeepCPS.
This can be encapsulated within an action of a grammar.
For example, when LangDSL defines itself, the rule for the semantic action (see \autoref{code:action}) is defined similarly to the code in \autoref{code:impl-foreign}.
\begin{code}
\begin{lstlisting}
Action|->(action)| ::= 
  ParamList|->(p_in)| ""->"" ParamList|->(p_out)|	""{""
	|(parin)->(head,lang)| ▼{
	  TRCreateLambdaHead parin (head)
		getCurrentLanguage ▲parser (lang)
		setCurrentLanguage ▲parser ▲DeepCPS ()
		return head lang
	}▼
	|(head)->|!LambdaBody|->(lambda)|
	|(lang,lambda,p_in,p_out)->(action)| ▼{
	  setCurrentLanguage ▲parser lang ()
		LangDSLCreateAction p_in p_out lambda return
	}▼
	""}"";
\end{lstlisting}
\caption{\label{code:impl-foreign}
A LangDSL semantic action construct defined in LangDSL itself.
The grammar calls a foreign production \inline{LambdaBody} of the \inline{DeepCPS} language.
}
\end{code}

In the above production, the first action invokes C functions to switch language from LangDSL to DeepCPS.
A nonterminal \inline{LambdaBody} from within DeepCPS is invoked.
The \inline{!} indicates a foreign nonterminal, which does not exist within the language where it is used.
When \inline{LambdaBody} has completed, the second action switches the currently parsed language back to LangDSL.

\section{Discussion}
\label{sec:discussion}

We have shown a unique compiler system that allows creation of multiple DSL with custom syntax,
yet embedding it in the functional language DeepCPS to express its semantics.
Our embedded DSLs are not limited by the syntax of the host language.
The semantic actions of a DSL can be used to build code directly without any additional intermediate representation.

The grammar specification itself is embedded in DeepCPS and is entirely function-based.
We have shown how portions of a grammar can be abstracted, creating reusable and parametrizable fragments of a grammar.

The flexibility and staging of DeepCPS allows for creating arbitrary control flow and introducing multiple passes for the target DSL.
The additional stages can be used to program domain-specific optimizations.
Staging can also be used to perform early program checks, implementing a custom type system or adding auxiliary computations~\cite{AuxiliaryComp}.

The code we generate is represented with fragment functions.
These fragments can be connected together akin to AST nodes.
Unlike an AST however, the functions are opaque: their behavior cannot be inadvertently changed in any way, by other nodes or tree transformations.
The only operations possible on a fragment functions is merging and execution.
Still, the execution may trigger optimization that is defined within the fragments through Dynamic Staging.

The opaqueness of the fragment functions gives each created language a unique possibility to define its own Language Programming Interface -- a set of entry rules that other languages may use. The LPI of the language defines entirely how it can be used and what can be produced with it.
The user does not need to worry about the internals of that language.

\subsection*{Future Work}

Within the paper, as well as in ManyDSL, we limit ourselves to LL(1) grammars.
Our main focus was to introduce the SDE scheme and language switching.
We plan to explore if and when this constraint can be lifted.
While SDE requires top-down parsing, supporting LL(*) or PEG is a possibility as long as backtracking is limited or avoided and language switching is suitably handled.
Alternatively, our LL(1) parser could be extended to support productions predicated by arbitrary DeepCPS code.

We continue to search for a good set of grammar abstractions within LangDSL.
We hope that nearly all aspects of DSL building can be deferred to a few grammar-building function calls.
Only the most unique syntactic constructs, specific for given domain, would require a direct grammar description.

Moreover, the handling of rule parameters in the abstraction is less than ideal.
We hope to find a more robust solution in the future.

Currently, the grammar actions can be expressed only directly in DeepCPS.
However, any language embedded in ManyDSL is suitable.
We want to increase productivity of LangDSL by permitting higher-level languages define the semantic actions.

In \autoref{ch:challenge:type} we described how Dynamic Staging can be used to define auxiliary computation and a simple type system.
In theory, nearly any type system can be defined as a staged computation and used in a custom DSL.
To our knowledge, however, this possibility has not yet been fully explored and require further research.

\printbibliography

\begin{code*}[H]
{\Large \textbf{Appendix}}
\vspace{3mm}
\begin{lstlisting}
Graph|->(P)| ::= |()->(Decl,Def)| ▼{
							     build 1 (▼'ft'▼, end, cont) {
								   newEnv (env) cont ▼'ft'▼ env 1 end
							   } (decl)
							   build 1 (▼'ft'▼, env, end, cont) {
								   cont ▼'ft'▼ env [] end
							   } (def)
							   return decl def
						   }▼
	             |(Decl,Def)->|Vertex|->(Decl,Def)|
	             |(Decl,Def)->(G)| ▼{
	               merge Decl Def (Descr)
	               build 0 (▼'ft'▼, env, graph, end) {
		               ▼'@ft:'▼ end graph
		             } (Fend)
		             merge Descr Fend (Descr)
		             finalize Descr return
	             }▼;
|(Decl,Def)->|Graph|->(Decl,Def)| ::= lassoc<Vertex, "";"", |\epsilon|>;
|(Decl,Def)->|Vertex|->(Decl,Def)| ::= 
               Name|->(name)|
				       |(Decl,name)->(Decl)| ▼{
	  		         build 1 (▼'ft'▼, env, idx, end, cont) {
					         ▼"env.insert(name,idx)"▼ (env)
						       ▼"idx+1"▼ (idx)
						       cont ▼'ft'▼ env idx end
					       } (DeclNext)
					       merge Decl DeclNext return
				       }▼
				       (Def)|->(Def)| ▼{
	  		         build 1 (▼'ft'▼, env, graph, end, cont) {
						       cont ▼'ft'▼ env graph [] end   //$\text{\textrm{adjacent list starting empty}}$
					       } (DefNext)
					       merge Def DefNext return
				       }▼				
				       ""->""
               lassoc<Edge, "","", epsilon>
				       (Def)|->(Def)| ▼{
	  		         build 1 (▼'ft'▼, env, graph, adjacent, end, cont)▼'[bt]'▼ {
					         ▼'@ft:' "concat(graph,[adjacent])"▼ (graph)▼'[ft]'
						       '@bt:'▼ cont ▼'ft'▼ env graph end
					       } (DefNext)
					       merge Def DefNext return
				       }▼;
				
|(Def)->|Edge|->(Def)| ::=
               Name|->(name)|
				       (Def,name)->(Def) ▼{
	  		         build 1 (▼'ft'▼, env, graph, adjacent, end, cont) {
					         ▼"env.lookup(name)"▼ (idx)▼'[bt]'
						       '@ft:' "concat(adjacent,[idx])"▼ (adjacent)▼'[ft]'
						       '@bt:'▼ cont ▼'ft'▼ env graph adjacent end
					       } (DefNext)
					       merge Def DefNext return
				       };
\end{lstlisting}
\caption{\label{code:graph-full}
The complete grammar for a graph-describing language from \autoref{ch:circular}. Two series of fragment functions are created: \inline{Decl} and \inline{Def}.
In \inline{Decl}, each new vertex is given a new index and added to an environment. Within \inline{Def} it is assumed that all vertices are already given a number.
These fragment functions are then merged in such a way that all \inline{Decl}-s precede all \inline{Def}-s.
}
\end{code*}

\end{document}